\documentclass[aps,preprint,showpacs]{revtex4}
\usepackage{graphicx,amsmath,amssymb}
\newcommand{\myext}{eps}

\newcommand{\e}{{\rm e}}
\newcommand{\F}{{\mathcal F}}
\newcommand{\ext}{{\rm ext}}
\newcommand{\exc}{{\rm exc}}

\begin{document}


\title{Replica density functional study of \\
       one-dimensional hard core fluids in porous media}

\author{Hendrik Reich}
\affiliation{
    Institut f\"ur Theoretische Physik II,
    Heinrich-Heine-Universit\"at D\"usseldorf, Universit\"atsstra\ss e 1,
    D-40225 D\"usseldorf, Germany.}
\author{Matthias Schmidt\footnote{On leave from:  
    Institut f\"ur Theoretische Physik II,
    Heinrich-Heine-Universit\"at D\"usseldorf, Universit\"atsstra\ss e 1,
    D-40225 D\"usseldorf, Germany.}}
\affiliation{
    Soft Condensed Matter, 
    Debye Institute, Utrecht University, Princetonplein 5,
    3584 CC Utrecht, The Netherlands.
}

\begin{abstract}
  A binary quenched-annealed hard core mixture is considered in one
  dimension in order to model fluid adsorbates in narrow channels
  filled with a random matrix. Two different density functional
  approaches are employed to calculate adsorbate bulk properties and
  interface structure at matrix surfaces. The first approach uses
  Percus' functional for the annealed component and an explicit
  averaging over matrix configurations; this provides numerically
  exact results for the bulk partition coefficient and for
  inhomogeneous density profiles. The second approach is based on a
  quenched-annealed density functional whose results we find to
  approximate very well those of the former over the full range of
  possible densities.  Furthermore we give a derivation of the
  underlying replica density functional theory.
\end{abstract}
\date{26 January 2004}
\date{4 February 2004}
\date{10 February 2004}


\maketitle

\section{Introduction}
\label{SECintroduction}
The one-dimensional hard rod model \cite{tonks36} continuous to be an
invaluable test bed for theoretical work as it provides the
possibility to compare approximations to exact results.  Recent
examples of this strategy include investigations of depletion
interactions in binary mixtures where one of the components is viewed
as an agent that mediates an effective interaction between particles
of the other component \cite{lekkerkerker00,oversteegen02}, a model
colloid-polymer mixture \cite{brader01oned} where particles
representing polymers can freely penetrate, dynamical density
functional theory \cite{marinibettolomarconi99,penna03} concerned with
time-dependent transport phenomena, as well as a model porous medium
\cite{kaminsky93} of lines of random length accessible to the fluid
particles.  As concerns equilibrium statistical mechanics, Percus'
exact free energy functional for pure systems \cite{Percus76} and
(additive) mixtures of particles with different sizes \cite{percus82},
provides a framework to compute thermodynamics, density distributions,
and correlation functions in arbitrary inhomogeneous situations.

Fluids adsorbed in disordered matrices are often described in the
context of so-called quenched-annealed (QA) fluid mixtures, where
particles of the quenched component act as randomly distributed
obstacles exerting an external potential on particles of the annealed
component \cite{maddenglandt,givenstell}. The matrix particles are
distributed according to the Hamiltonian of the quenched component,
and hence can be treated with liquid state theory.  The crucial
difference to an equilibrium system of two annealed components is that
the distribution of matrix particles is unaffected by the presence of
the annealed component.  As one is interested in the typical behavior
of the system, a double average over the annealed degrees of freedom
and over the quenched disorder is required. A short overview of this
theoretical framework is given below. One standard approach to tackle
fluid structure and phase behavior of QA models is via the replica
Ornstein-Zernike relations \cite{maddenglandt,givenstell} supplemented
with appropriate closure relations. Many standard liquid integral
equation theories have been carried over to the QA case. The basic
quantities in terms of which these theories are formulated are
two-body (and possibly higher) correlation functions.

Recently it was proposed to rather work directly on the level of the
free energy functional, pre-averaged over the disorder
\cite{schmidt02pordf}. Correlation functions can be obtained
subsequently, in particular the hierarchy of direct correlation
functions is obtained through functional differentiation with respect
to the density fields. The advantage of formulating the theory on the
one-body level of the density fields is that inhomogeneous situations,
e.g.\ at free interfaces or caused by external fields like e.g.\
gravity, are straightforward to treat. Hence all advantages of
equilibrium DFT \cite{evans79,evans92} apply to this QA or replica
DFT. Also the disadvantage applies; in general the density functional
is unknown.  Moreover, whether one can learn anything about the
important out-of-equilibrium behavior of QA systems, like hysteresis
in sorption isotherms, is questionable. Based on Rosenfeld's
fundamental-measure theory (FMT) for hard sphere mixtures
\cite{rosenfeld89}, and the subsequent discovery that one can
construct density functionals by imposing the correct behavior upon
dimensional reduction (i.e.\ in situations of extreme confinement in
one or more spatial directions through suitably chosen external
potentials) \cite{RSLTlong,tarazona97}, this DFT treats QA mixtures
with either hard or ideal interactions.

Previous tests of QA DFT include the comparison with results from
Monte Carlo (MC) simulations for the partial pair correlation
functions in hard sphere systems \cite{schmidt02pordf} and for density
profiles across the surface of a porous medium, modeled as a step-like
density distribution of (freely overlapping) matrix spheres
\cite{schmidt03porz}. Such an interface was also investigated in a
model of a random fiber network, being represented by quenched
configurations of infinitely thin needle-like particles. Again
comparison with simulation results shows satisfactory agreement with
DFT results \cite{schmidt03porsn}.  Much work has been devoted to a
model colloid-polymer mixture where the colloids are represented as
hard spheres and the polymers as freely overlapping spheres. The
crucial step beyond hard sphere systems is the occurrence of a
fluid-fluid phase transition, and the questions how capillary
condensation occurs inside a porous medium \cite{schmidt02aom} and
what the structure of the interface between demixed fluid phases is
like \cite{wessels03aoim} were treated.  A further, very promising,
line of research is the application of the approach to lattice
models. Note that using lattice models insight into hysteresis
behavior and the relation to the appearance of a complex free energy
landscape was gained \cite{kierlik01,kierlik02}.  Combining the QA-DFT
approach with the very powerful lattice DFT by Lafuente and Cuesta
\cite{lafuente02prl,lafuente02}, freezing in a two-dimensional
lattice model was investigated \cite{schmidt03lfmf}.

In this work we consider the one-dimensional hard rod model adsorbed
in a quenched matrix of rods. Two cases of interactions between the
quenched particles are considered.  In the first case the rods
interact with a hard core potential, hence we deal with a binary QA
hard rod mixture. In the second case the matrix particle are ideal
(non-interacting) amongst each other, but interact with a hard-core
potential with particles of the annealed component; this model is the
QA analog of the model colloid-polymer mixture of Ref.\
\cite{brader01oned}, obtained by quenching the polymers. We use two
different density functional approaches to tackle the properties of
these models in both homogeneous and inhomogeneous (on average over
disorder) situations. In the first approach the matrix is treated
explicitly on the level of particle coordinates distributed according
to the matrix Hamiltonian, and practically generated with a Monte
Carlo procedure. For each matrix configuration we use Percus'
functional \cite{Percus76} to obtain adsorbate properties, and the
disorder-average over matrix configuration is carried out numerically
by brute force generation of many (of the order of 1000) matrix
realizations.  The second approach is the QA DFT working directly on
the level of the disorder-averaged adsorbate density profile, which is
obtained via minimizing the disorder-averaged grand potential of the
adsorbate, using the matrix density profiles as a fixed input. As the
average over disorder is taken a priori, the corresponding
Euler-Lagrange equation yields directly the averaged density profile.
We compare results from both theories in bulk via calculation of the
partition coefficient, which is the ratio of adsorbate density inside
the matrix and the density in a bulk reservoir that is in chemical
equilibrium. The results from the explicit matrix averaging procedure,
which we also check against an independent elementary calculation,
agree well with those from the QA DFT over the full range of
accessible densities. Deviations appear at high densities, which we
can trace back to an incorrect behavior of the QA-DFT near
close-packing.  As a generic inhomogeneous situation we consider a
surface of the model porous matrix, that is generated by a hard wall
acting on the matrix particles (before the quench). The wall is then
removed and leaves a halfspace of bulk (free of matrix particles). The
adsorbate is found to exhibit density oscillations on both sides of
the interface, with significantly smaller amplitude than at a hard
wall.

The paper is organized as follows. In Sec.\ \ref{SECmodel} we define
the model and give an overview of its statistical mechanics and the
replica trick (which can be safely skipped by an expert reader). Sec.\
\ref{SECdft} is devoted to both density functional methods. In Sec.\
\ref{SECresults} results are presented and we conclude in Sec.\
\ref{SECconlusions}.

\section{The model}
\label{SECmodel}
\subsection{Definition of the interactions}
\begin{figure}
\includegraphics[width=0.6\columnwidth]{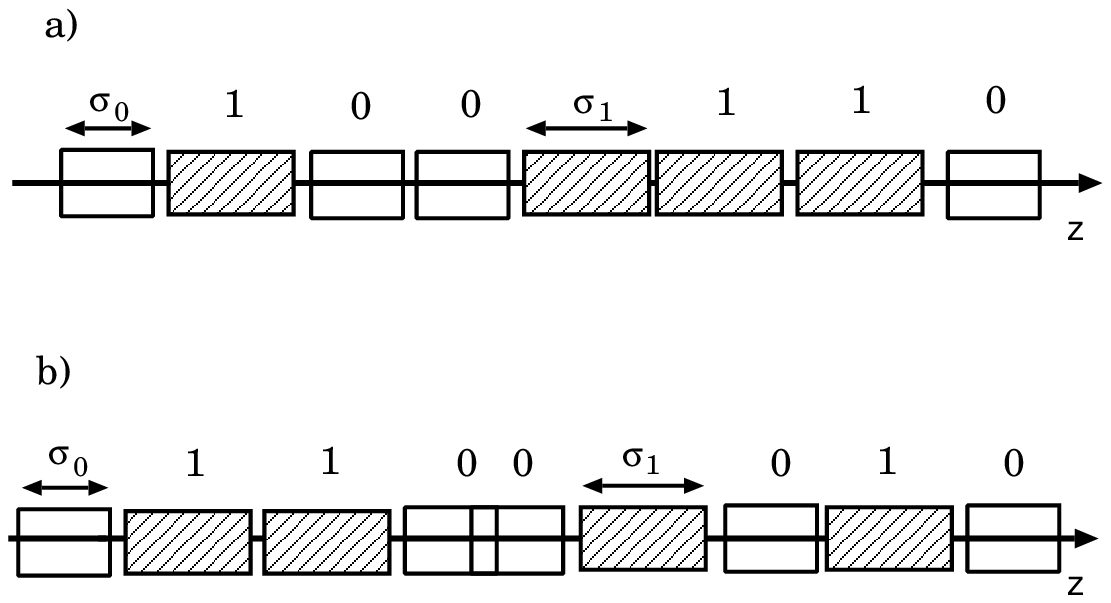}
\caption{Model of annealed hard rods in a matrix of quenched rods in
  one dimension. The quenched rods interact with a) a hard core
  potential, and b) are ideal.}
\label{FIGmodel}
\end{figure}

We consider a quenched-annealed fluid mixture of a quenched species 0
with $N_0$ particles with one-dimensional (1d) position coordinates
$x_1,x_2,\ldots x_{N_0}$ and an annealed species 1 with $N_1$
particles with 1d position coordinates $X_1,X_2,\ldots X_{N_1}$. The
particles interact with pairwise potentials 
(for pairs $\alpha\gamma=00,01,11$) 
given by
\begin{equation}
 \phi_{\alpha\gamma}(x) = 
  \begin{cases}\infty & x < (\sigma_\alpha+\sigma_\gamma)/2\\
   0 & \rm otherwise, \end{cases}
  \label{EQphiAlphaGamma}
\end{equation}
where $x$ is the center-center distance between two particles;
$\sigma_\alpha$ is the diameter (length) of particles of species
$\alpha=0,1$, see Fig.\ \ref{FIGmodel}a for an illustration.
This describes our first case of the hard core matrix. The total
potential energy due to particle-particle interactions is
$V_{00}+V_{01}+V_{11}$, with contributions
\begin{eqnarray}
 V_{00} &=& \sum_{i=1}^{N_0}\sum_{j=i+1}^{N_0} \phi_{00}(|x_i-x_j|),
 \label{EQvZeroZero}\\
 V_{01} &=& \sum_{i=1}^{N_0}\sum_{j=1}^{N_1}   \phi_{01}(|x_i-X_j|),
 \label{EQvZeroOne}\\
 V_{11} &=& \sum_{i=1}^{N_1}\sum_{j=i+1}^{N_1} \phi_{11}(|X_i-X_j|).
 \label{EQvOneOne}
\end{eqnarray}
We furthermore consider the influence of external potentials,
$\phi_\alpha^\ext(x)$, acting on species $\alpha=0,1$,
respectively. In particular $\phi_0^\ext(x)$ acts on particles of
species 0 {\em before} they are quenched, i.e.\ their density
distribution is that generated in response to $\phi_0^{\rm
ext}(x)$. The resulting total external potential energy is $V_0^{\rm
ext}+V_1^\ext$, with contributions
\begin{equation}
  V_0^\ext = \sum_{i=1}^{N_0} \phi_0^\ext(x_i),\quad
  V_1^\ext = \sum_{i=1}^{N_1} \phi_1^\ext(X_i).
\end{equation}
In our second case we consider ideal (non-interacting) matrix
particles, i.e.\
\begin{equation}
 \phi_{00}(x)=0,\label{EQphi00id}
\end{equation}
valid for all distances $x$;
the two remaining interactions are unchanged, i.e.\ $\phi_{01}(x)$ and
$\phi_{11}(x)$ are hard core potentials given through
\eqref{EQphiAlphaGamma};
see Fig.\ \ref{FIGmodel}b for an illustration. The size ratio
$s=\sigma_1/\sigma_0$ is a geometric control parameter.  In the
numerical results presented below we will restrict ourselves to
equally sized particles, $s=1$, and furthermore to situations where
$\phi_1^\ext(x)=0$.

\subsection{Partition sum, grand potential and replica trick}
We first make the statistical mechanics of the quenched-annealed
mixture explicit. For notational convenience the grand canonical trace
over matrix coordinates is denoted by $\int
d0\equiv\sum_{N_0=0}^\infty \left(N_0!\Lambda_0^{N_0}\right)^{-1} \int
dx_1 \ldots \int dx_{N_0}$ and that over adsorbate coordinates by
$\int d1\equiv\sum_{N_1=0}^\infty
\left(N_1!\Lambda_1^{N_1}\right)^{-1} \int dX_1 \ldots \int dX_{N_1}$,
where $\Lambda_\alpha$ is the (irrelevant) thermal wavelength of
species $\alpha=0,1$, and the position integrals run over the total
system volume $V$.  The (equilibrium) grand partition of the matrix
particles under the influence of the external potential
$\phi_0^\ext(x)$ is
\begin{equation}
 \Xi_0(\mu_0,T,V) = \int d0 \e^{-\beta(V_{00}+V_0^\ext-\mu_0 N_0)},
\end{equation}
where $\beta=1/(k_BT)$, $k_B$ is the Boltzmann constant, $T$ is
absolute temperature and $\mu_i$ is the chemical potential of species $i=0,1$. The grand potential for the matrix is then
\begin{equation}
\Omega_0(\mu_0,T,V) = -\beta^{-1}\ln \Xi_0(\mu_0,T,V).
\end{equation}
For fixed matrix configuration $\{x_i\}$ the grand potential of the
adsorbate is
\begin{equation}
  \Omega_1(\{x_i\},\mu_1,T,V) = 
   -\beta^{-1} \ln \int d1 \e^{-\beta(V_{11}+V_{01}+V_1^\ext-\mu_1N_1)},
   \label{EQomegaOneFixedZeroConfiguration}
\end{equation}
depending explicitly on $\{x_i\}$ through $V_{01}$, see
\eqref{EQvZeroOne}. Note that from the viewpoint of the 1-particles,
$V_{01}+V_1^\ext$ is the {\em total} external potential energy. A
(grand canonical) average over matrix configurations yields the
disorder-averaged grand potential of the adsorbate,
\begin{equation}
  \Omega_1(\mu_0,\mu_1,T,V) = \Xi_0^{-1}(\mu_0,T,V) \int
  d0\e^{-\beta(V_{00}+V_0^\ext-\mu_0N_0)} \Omega_1(\{x_i\},\mu_1,T,V).
  \label{EQomegaOneExplicitAverage}
\end{equation}
Note that \eqref{EQomegaOneExplicitAverage} has a different structure
than that of the partition sum of an equilibrium mixture due to the
appearance of the logarithm, upon inserting
\eqref{EQomegaOneFixedZeroConfiguration} into
\eqref{EQomegaOneExplicitAverage}, {\em inside} the trace over the
0-particles.  However, a relation to a multi-component mixture can be
established using the replica trick: One introduces replicas as $s$
copies of species 1: $\phi_{\alpha\alpha}(x)=\phi_{11}(x)$, for
$1<\alpha\leq s$, where $s$ is an integer. Particles from different
replicas are non-interacting, $\phi_{\alpha\gamma}(x)=0$ for all $x$
and $\alpha\neq\gamma$, but they interact with matrix particles in the
same fashion, $\phi_{0\alpha}(x)=\phi_{01}(x)$, for $1<\alpha\leq
s$. Then the (equilibrium) partition sum for this $(s+1)$-component
mixture can be written as
\begin{eqnarray}
 \Xi &=& \int d0 \e^{-\beta(V_{00}+V_0^\ext-\mu_0N_0)} 
 \left(\int d1  \e^{-\beta(V_{11}+V_{01}+V_1^\ext-\mu_1N_1)}
 \right)^s,
\end{eqnarray}
and the grand potential is
\begin{equation}
  \Omega(\mu_0,\mu_1,T,V;s) = -\beta^{-1}\ln\Xi.
\end{equation}
Via analytical continuation in $s$, and noting that $\lim_{s\to
0}dx^s/ds=\ln x$, the disorder-averaged grand potential,
\eqref{EQomegaOneExplicitAverage}, is obtained from the equilibrium
grand potential of the replicated system as
\begin{equation}
 \Omega_1(\mu_0,\mu_1,T,V) = \lim_{s\to 0}
 \frac{d}{ds}\Omega(\mu_0,\mu_1,T,V;s),
 \label{EQomegaOneFromReplicatedModel}
\end{equation}
establishing a practical route to tackle the QA system via the
replicated equilibrium system.

\section{Density functional approaches}
\label{SECdft}
The following subsections \ref{SECequilibriumDFT},
\ref{SECquenchedAnnealedDFT} are valid in arbitrary space dimension
$d$ upon trivial alterations: spatial integrations become
$d$-dimensional integrals, hence $\int dx$ is to be replaced with
$\int d^dx$ and factors $\Lambda_\alpha$ are to be replaced with
$\Lambda_\alpha^d$.  Although in our subsequent study we only deal
with hard core interactions, where the dependence on temperature is
trivial, the formalism also applies to thermal systems.  Hence we
consider a general binary QA mixture with arbitrary pair potentials
$\phi_{\alpha\gamma}(x)$, $\alpha,\gamma=0,1$, not necessarily given
through 
\eqref{EQphiAlphaGamma}, at temperature $T$ inside a volume $V$.

\subsection{Equilibrium case}
\label{SECequilibriumDFT}
In the equilibrium DFT formalism, applied to the present case, where
both an explicit external potential, $\phi_1(x)$, and the (random)
influence of the matrix particles at positions $\{x_i\}$ acts on the
fluid, the grand potential of the adsorbate component is expressed as
a functional of its one-body density distribution,
\begin{eqnarray}
 \tilde\Omega_1(\{x_i\},[\rho_1],\mu_1,T,V) = &&
   \F^{\rm id}[\rho_1] + \F^\exc[\rho_1]\nonumber\\&&
   + \int dx \rho_1(x) \left[
   \left(\phi_1^\ext(x)+\sum_{i=1}^{N_0}
   \phi_{01}(x-x_i)\right)-\mu_1\right],
   \label{EQomegaOneEquilibrium}
\end{eqnarray}
where $\F^{\rm id}[\rho_1]=\beta^{-1}\int dx
\rho_1(x)[\ln(\rho_1(x)\Lambda_1)-1]$ is the (Helmholtz) free energy
functional of the ideal gas, $\F^\exc[\rho_1]$ is the excess (over
ideal) contribution that arises from interactions between (adsorbate)
particles, and the term in round brackets is the {\em total} external
potential acting on the adsorbate stemming from the explicit
(non-random) external potential, $\phi_1^\ext(x)$, and the sum over
interactions with (randomly distributed) matrix particles. The latter
contribution is parameterized by the set of matrix coordinates
$\{x_i\}$, and hence $\tilde\Omega_1$ depends explicitly on $\{x_i\}$,
which we stress in the notation of the l.h.s.\ of
\eqref{EQomegaOneEquilibrium}.  The free energy functionals $\F^{\rm
id}$ and $\F^\exc$ 
depend on $T$ and $V$; this is suppressed in the notation in
\eqref{EQomegaOneEquilibrium} and in the following.  The minimization
condition is
\begin{equation}
 \left.\frac{\delta \tilde\Omega(\{x_i\},[\rho_1],\mu_1,T,V)}{\delta \rho_1}
 \right|_{\rho_1=\rho_1(\{x_i\},x)} = 0,
 \label{EQomegaOneMinimizedExplicitMatrix}
\end{equation}
where $\rho_1(\{x_i\},x)$ is the adsorbate density distribution that
solves \eqref{EQomegaOneMinimizedExplicitMatrix}.  The value of the
grand potential is then obtained by reinserting the solution into the
grand potential functional,
\begin{equation}
 \Omega(\{x_i\},\mu_1,T,V) = \tilde\Omega(\{x_i\},[\rho_1(x,\{x_i\})],\mu_1,T,V),
\end{equation}
from which the average over the disorder, $\Omega(\mu_0,\mu_1,T,V)$,
can be obtained via \eqref{EQomegaOneExplicitAverage}.  In a similar
way as for the grand potential, the matrix-averaged adsorbate density
profile is given as
\begin{equation}
 \rho_1(x) = \Xi_0^{-1}(\mu_0,T,V) \int d0 
 \e^{-\beta(V_{00}+V_0^\ext-\mu_0N_0)} \rho_1(x,\{x_i\}).
 \label{EQrhozAverage}
\end{equation}
Note that {\em explicit} averages over the matrix configurations are
to be performed in \eqref{EQomegaOneExplicitAverage} and
\eqref{EQrhozAverage}. In the numerical procedure described below, we
will carry this out numerically via a Monte Carlo procedure.

\subsection{Quenched-annealed case}
\label{SECquenchedAnnealedDFT}
Here we first formulate the equilibrium DFT of the replicated model
from which we obtain, in the appropriate limit of vanishing number of
components, the minimization condition of DFT for one quenched and one
annealed component.  Explicitly assuming absence of replica symmetry
breaking, hence $\rho_1(x)=\rho_\alpha(x), 1<\alpha\leq s$, the grand
potential functional for the replicated {\em equilibrium} mixture
reduces to
\begin{equation}
  \tilde\Omega([\rho_0,\rho_1],\mu_0,\mu_1,T,V;s) \equiv
  \tilde\Omega([\{\rho_\alpha\}],\{\mu_\alpha\},T,V)
  \label{EQomegaMix}
\end{equation}
The minimization conditions are
\begin{eqnarray}
 \frac{\delta \tilde\Omega([\rho_0,\rho_1],\mu_0,\mu_1,T,V;s)}{\delta
 \rho_\alpha(x)} &=& 0, \quad \alpha=0,1,
 \label{EQminMultiComponent}
\end{eqnarray}
which are two {\em coupled} equations for the two unknown functions
$\rho_0(x)$ and $\rho_1(x)$.  At the minimum the value of the
functional is the true grand potential
\begin{eqnarray}
  \Omega(\mu_0,\mu_1,V,T;s) = \tilde\Omega([\rho_0,\rho_1],\mu_0,\mu_1,T,V;s).
\end{eqnarray}
Via analytic continuation, and Taylor expanding in s around $s=0$, one
obtains
\begin{equation}
 \tilde\Omega([\rho_0,\rho_1],\mu_0,\mu_1,T,V;s) = 
 \tilde\Omega_0([\rho_0],\mu_0,T,V) + 
 s\tilde\Omega_1([\rho_0,\rho_1],\mu_1,T,V)+O(s^2),
 \label{EQomegaSmallsExpansion}
\end{equation}
where $\tilde\Omega_0$ is the grand potential of the pure system of
0-particles, that may formally be written as
$\tilde\Omega_0([\rho_0],\mu_0,T,V)=
\tilde\Omega([\rho_0,0],\mu_0,\mu_1\to-\infty,T,V)$, furthermore
$\tilde\Omega_1[\rho_0,\rho_1] = \lim_{s\to 0}
d\tilde\Omega([\rho_0,\rho_1],\mu_0,\mu_1,T,V;s)/ds$. Note that via
this definition $\tilde\Omega_1$ is also the disorder-averaged grand
potential, given in \eqref{EQomegaOneFromReplicatedModel}.

We decompose both contributions in the standard way:
\begin{eqnarray}
 \tilde\Omega_0([\rho_0],\mu_0,T,V) &=& \F^{\rm id}[\rho_0] + \F_0^\exc[\rho_0] +
  \int dx \rho_0(x)\left(\phi_0^\ext(x) - \mu_0\right),
  \label{EQomegaZeroQA}\\
  \tilde\Omega_1([\rho_0,\rho_1],\mu_1,T,V) &=& 
  \F^{\rm id}[\rho_1] + \F_1^{\rm exc}[\rho_0,\rho_1] + 
  \int dx \rho_1(x)\left(\phi_1^\ext(x) - \mu_1\right),
  \label{EQomegaOneQA}
\end{eqnarray}
which can be viewed as definitions for the excess (over ideal gas)
free energy functionals, $\F_0^\exc[\rho_0]$ and
$\F_1^\exc[\rho_0,\rho_1]$, that arise from interactions between
particles. Note, however, that neither $\F^{\rm id}[\rho_0]$ nor a
contribution involving $\phi_0^{\rm ext}(x)$ appear on the r.h.s.\ of
\eqref{EQomegaOneQA}, in contrast to the binary equilibrium case.

We insert the small-$s$-expansion of the grand potential functional,
\eqref{EQomegaSmallsExpansion}, into the equilibrium minimization
condition, \eqref{EQminMultiComponent}. Performing the limit $s\to 0$,
the minimization conditions for the QA mixture result as
\begin{eqnarray}
 \frac{\delta\tilde\Omega_0([\rho_0],\mu_0,T,V)}
      {\delta\rho_0(x)}&=&0,\label{EQminZero}\\
 \frac{\delta\tilde\Omega_1([\rho_0,\rho_1],\mu_1,T,V)}
      {\delta\rho_1(x)}&=&0\label{EQminOne}.
\end{eqnarray}
The derivation of \eqref{EQminZero} is straightforward. To obtain
\eqref{EQminOne} one sets $\alpha=1$ in \eqref{EQminMultiComponent},
i.e.\ differentiates w.r.t.\ to $\rho_1(x)$ and divides the resulting
equation by $s$, assuming $s>0$, before taking the limit $s\to 0$.
Note that \eqref{EQminZero} is decoupled from \eqref{EQminOne}, hence
in particular $\rho_0(x)$ is solely determined through
\eqref{EQminZero}. The result then serves as an input to
\eqref{EQminOne}, which is solely to be solved for $\rho_1(x)$.

\subsection{Excess free energy functionals}
The theoretical approaches described so far incorporate the complexity
of the problem i) in the excess free energy functionals for the
adsorbate, $\F^\exc[\rho_1]$, and an explicit average over matrix
configurations, and ii) in excess free energy functionals for the
matrix component, $\F_0^\exc[\rho_0]$, and for the adsorbate in the
presence of the matrix, $\F_1^\exc[\rho_0,\rho_1]$. For the present 1d
model, the situations is fortunate, as an exact result for
$\F^\exc[\rho_1]$ and $\F_0^\exc[\rho_0]$ is available, namely Percus'
free energy functional for 1d hard rods \cite{Percus76,percus82}.  For
a one-component system of hard rods of species $\alpha$ one writes
(using Rosenfeld's terminology)
\begin{equation}
  \F^\exc[\rho_\alpha] = 
  \int dx \, n_\alpha^{(0)}(x) \Phi'_{\rm hc}\left(n_\alpha^{(1)}(x)\right),
  \label{EQfexcOneComponent}
\end{equation}
where the weighted densities, $n_\alpha^{(0)}(x)$ and
$n_\alpha^{(1)}(x)$, are obtained from the bare density profile (of
species~$\alpha$) via
\begin{eqnarray}
  n_\alpha^{(0)}(x) &=& [\rho_\alpha(x-R_\alpha)+\rho_\alpha(x+R_\alpha)]/2, 
  \label{EQnZero}\\
  n_\alpha^{(1)}(x) &=& \int_{x-R_\alpha}^{x+R_\alpha} 
  dx' \rho_\alpha(x') \label{EQnOne},
\end{eqnarray}
where $R_\alpha=\sigma_\alpha/2$ is the particle ``radius'' of species
$\alpha=0,1$, and the upper index $\nu=0,1$ of the weighted density is
related to its dimension, which is $({\rm length})^{\nu-d}$, where
$d=1$ is the space dimension. The prime in \eqref{EQfexcOneComponent}
denotes differentiation w.r.t.\ the argument, and $\Phi_{\rm
hc}(\eta)$ is the zero-dimensional free energy of hard core particles
\cite{RSLTlong}, given by
\begin{equation}
 \Phi_{\rm hc}(\eta) = (1-\eta)\ln(1-\eta)+\eta.
 \label{EQphiZeroDimHC}
\end{equation}

Our (approximate) QA functional has very similar structure
\cite{schmidt02pordf}. We start from the generalization of
\eqref{EQfexcOneComponent} to binary mixtures, which is
\begin{equation}
  \F_1^\exc[\rho_0,\rho_1] = 
  \int dx \sum_{\alpha=0,1} n_\alpha^{(0)}(x)
  \Phi_\alpha\left(n_0^{(1)}(x), n_1^{(1)}(x)\right), \label{EQfexcTwoComponent}
\end{equation}
where the weighted densities are still given through \eqref{EQnZero}
and \eqref{EQnOne}, and derivatives of the zero-dimensional free
energy, $\Phi$, are defined as $\Phi_\alpha(\eta_0,\eta_1) = \partial
\Phi(\eta_0,\eta_1)/\partial \eta_\alpha$.
The particular form \eqref{EQfexcTwoComponent} ensures that the exact
result for $\Phi$ is recovered if the functional is applied to a
zero-dimensional density distribution, defined as $\rho_i(x)=\eta_i
\delta(x)$ for $i=0,1$. Hence it can be shown via elementary
calculation that $\Phi(\eta_0,\eta_1)=\F_1^\exc[\eta_0 \delta(x),
\eta_1 \delta(x)]$.
The exact result for $\Phi(\eta_0,\eta_1)$ is obtained by solving the
zero-dimensional limit (where all particles present in the system
overlap); a detailed calculation can be found in Ref.\
\cite{schmidt02pordf}.
Such a situation can be enforced by appropriate external potentials
$\phi_\alpha^\ext(x)=\infty$ if $|x|>L$ and zero otherwise, and
represents a ``cavity'' of size $L$, where
$L\ll\min(\sigma_0,\sigma_1)/2$.  Hence any two particles (of species
$\alpha$ and $\gamma$) present in the system will overlap, i.e.\
$x<\sigma_{\alpha\gamma}$, where $x$ is the center-center distance
between both particles. (As a consequence the density profiles vanish
outside the cavity, $\rho_\alpha(|x|>L/2)=0$.) This constitutes the
crucial simplification that allows to calculate the free energy; its
excess contribution is independent of $L$ \cite{schmidt02pordf}.
In the case of the hard core matrix the result is
\begin{equation}
  \Phi(\eta_0,\eta_1) = (1-\eta_0-\eta_1)\ln(1-\eta_0-\eta_1)+\eta_1 -
  (1-\eta_0)\ln(1-\eta_0).
\end{equation}
It is interesting to note that in this special case there is a simple
relation to the corresponding fully annealed binary hard core mixture:
$\Phi(\eta_0,\eta_1)=\Phi_{\rm hc}(\eta_0+\eta_1)-\Phi_{\rm
hc}(\eta_0)$. In the case of the ideal matrix the result for the
excess free energy is
\begin{equation}
  \Phi(\eta_0,\eta_1) = 
  \left(\e^{-\eta_0}-\eta_1\right)\ln\left(\e^{-\eta_0}-\eta_1\right)+\eta_1
  +\eta_0\e^{-\eta_0}.
\end{equation}
This does not have a similar relation as above to the corresponding
fully annealed binary mixture (the colloid-polymer mixture of Ref.\
\cite{brader01oned}).  As a further aside, we note that setting
$\Phi(\eta_0,\eta_1)=\Phi_{\rm hc}(\eta_0+\eta_1)$ in
\eqref{EQfexcTwoComponent} gives the exact excess free energy
functional for equilibrium (where both species are annealed) binary
hard rods \cite{Percus76,percus82}.

The density functional thus defined is exact on the second virial
level, which can be seen by Taylor expanding $\Phi(\eta_0,\eta_1)$ in
both arguments to second order, and exploiting the property of the
weight functions to recover the Mayer bond upon convolution.

\subsection{Numerical procedure}
In order to calculate density profiles, the minimization is done with
a standard iteration technique. We chose a very fine grid with spacing
$0.0002\sigma$ to discretize the space coordinate. The system is
assumed to be periodic in $x$ with length $20\sigma$.  To generate the
matrix configurations $\{x_i\}$, a Monte Carlo scheme is used. We
start from an initial configuration where the (matrix) particles have
equally-spaced positions. Then particle displacements are performed
according to the Metropolis algorithm, i.e.\ a new position is only
accepted if no overlap with any other (matrix) particle or with the
wall (introduced below) occurs. 1000 MC moves per particle are performed for
equilibration, and 100 MC moves per particle are performed between
configurations that are used for data production.  For given matrix
configuration $\{x_i\}$ the density profile for the adsorbate
component, $\rho_1(x,\{x_i\})$, is then obtained from solution of the
minimization condition of the equilibrium DFT,
\eqref{EQomegaOneMinimizedExplicitMatrix}. Results from 1000
independent matrix realizations for each $N_0$ in the range $N_1=0-20$
are then used to carry out the average over the disorder and obtain
$\rho_1(x)$ via \eqref{EQrhozAverage}.

\section{Results}
\label{SECresults}

\subsection{Bulk}
The bulk case constitutes the basis of the subsequent interface study,
and is considered for the following three reasons: i) to assess the
accuracy of the QA-DFT, ii) to demonstrate the correctness of the
matrix-averaging procedure via comparing with an independent
elementary calculation, and iii) to study the (exact) partition
coefficient, which we find to possess (very unusual) non-monotonic
behavior.

The matrix particles are distributed homogeneously on average, and
hence are characterized by a one-body density distribution
$\rho_0(x)=\eta_0/\sigma_0=\rm const$. As a consequence, the adsorbate
density distribution is on average $\rho_1(x)=\eta_1/\sigma_1=\rm
const$. We imagine the system to be in chemical equilibrium with a
reservoir of hard rods of species 1 of packing fraction
$\eta_1^r=\rho_1^r\sigma_1$, where $\rho_1^r$ is the number density in
the reservoir; there are no quenched matrix particles in the
reservoir. The reservoir density sets the chemical potential of
1-particles via the (well-known) hard rod equation of state,
\begin{equation}
\beta\mu_1 = \ln(\eta_1^r\Lambda_1/\sigma_1)
-\ln(1-\eta_1^r)+\frac{\eta_1^r}{1-\eta_1^r}.
\label{EQreservoirEqos}
\end{equation}
The central quantity that we use to characterize the coupled systems
is the partition coefficient $K=\eta_1/\eta_1^r=\rho_1/\rho_1^r$,
which we will study as a function of the reservoir packing
fraction~$\eta_1^r$.

Before applying both DFT methods to this case, we first seek to obtain
a benchmark result for $K(\eta_1^r)$ from an independent, elementary
calculation; see \cite{torquato90} for more mathematical background
and \cite{kaminsky93} for an alternative application.  We start from
the probability $W(x)dx$ that the nearest-neighbor distance for
0-particles (measured between their centers) is between $x$ and $x+dx$
(we call this a gap of size $x$), given by
\begin{equation}
 W(x)\sigma_0 = \begin{cases} 
   \frac{\eta_0}{1-\eta_0}
   \exp\left(\frac{\eta_0}{1-\eta_0}(1-x/\sigma_0)\right) &
   x>\sigma_0\\
   0 & \rm otherwise, \end{cases}
 \label{EQwofx}
\end{equation}
from which the average nearest-neighbor distance between matrix
particles follows as $\bar x=\int dx xW(x)=\sigma_0/\eta_0$, which is
an expected result.  To derive \eqref{EQwofx}, note that the
occurrence of a gap of size $x$ is proportional to the Boltzmann
weight of the reversible mechanical work to create it, hence
$W(x)\propto\exp(-P_0x)$, where $P_0$ is the pressure of the bulk
(matrix) system, given for one-dimensional hard rods by $\sigma_0\beta
P_0 = \eta_0/(1-\eta_0)$. One obtains the precise form of
\eqref{EQwofx} by taking into account the correct normalization,
$\int_0^\infty dxW(x)=1$.

The average number of 1-particles in a gap of (fixed) size $x$ between
two 0-particles is $\bar N_1(x,\mu_1)=\beta^{-1}\partial \ln
\Xi_1/\partial\mu_1$, where the partition sum of 1-particles in the
gap of volume $x-\sigma_0-\sigma_1$ (being accessible to the centers
of 1-particles) is
\begin{equation}
  \Xi_1(x,\mu_1) = \int d1 \exp(-\beta(V_{11}-\mu_1N_1))=
  \sum_{N_1=0}^\infty 
  \frac{\exp(\beta\mu_1N_1)}{\Lambda_1^{N_1}N_1!} 
  (x-\sigma_0-N_1\sigma_1)^{N_1}.
\end{equation}
The mean number of 1-particles in the gap between two neighboring
0-particles, averaged over all sizes of the gap, is $ \bar N_1(\mu_1)
= \int dx W(x) \bar N_1(x,\mu_1)$.  Then the average density of
1-particles is $\rho_1 = \bar N_1(\mu_1)/\bar x = \bar
N_1(\mu_1)\eta_0/\sigma_0$, from which the partition coefficient
results as $K = \rho_1/\rho_1^r=\bar N_1\sigma_1/(\bar
x\eta_1^r)$. Putting things together,
\begin{equation}
  K = \frac{\eta_0\sigma_1}{\eta_1^r\sigma_0} \int_0^\infty dx W(x) 
  \frac{\partial \ln\Xi_1(x,\mu_1)}{\partial \beta\mu_1},
  \label{EQkFinal}
\end{equation}
where the equation of state in the reservoir of 1-particles,
\eqref{EQreservoirEqos}, can be used to obtain $\mu_1$ in terms of
$\eta_1^r$ in \eqref{EQkFinal}, hence $K$ is solely a function of the
packing fractions $\eta_0, \eta_1^r$ and of the size ratio
$\sigma_1/\sigma_0$. In general, we solve \eqref{EQkFinal}
numerically; we can, however, obtain analytic results in both
(extreme) cases of high and low reservoir density,
\begin{eqnarray}
 K(\eta_1^r\to 0) &=& (1-\eta_0) \exp(-s\eta_0/(1-\eta_0)),
 \label{EQkHardCoreLowDensity}\\
 K(\eta_1^r\to 1) &=&
 \frac{s\eta_0}{\exp\left(s\eta_0/(1-\eta_0)\right)-1},
 \label{EQkHardCoreHighDensity}
\end{eqnarray}
where $s=\sigma_1/\sigma_0$.  \eqref{EQkHardCoreHighDensity} is
obtained by noting that for $\mu_1\to\infty$ the mean number of
1-particles equals the maximal possible number, i.e.\ $\bar
N_1(x,\mu_1\to\infty) = {\rm floor}((x-\sigma_0)/\sigma_1)$, where
${\rm floor}(x)$ gives the next integer smaller than $x$.

In the case of the of the non-interacting matrix, the equation of
state of the 0-particles is that of an ideal gas, $\sigma_0\beta
P=\eta_0$.  The gap size distribution is
\begin{equation}
 W(x) = (\eta_0/\sigma_0) \exp(-\eta_0 x/\sigma_0), \quad x>0,
\end{equation}
from which $K$ can be derived following the same steps as above.
Again in the two limiting cases of either high or low reservoir
packing fraction analytic expression can be obtained for the partition
coefficient,
\begin{eqnarray}
 K(\eta_1^r\to 0) &=& \exp(-(1+s)\eta_0), \label{EQkIdealLowDensity}\\
 K(\eta_1^r\to 1) &=& \frac{s\eta_0\exp(-\eta_0)}{\exp(s\eta_0)-1}.
 \label{EQkIdealHighDensity}
\end{eqnarray}

In order to obtain $K$ from the QA DFT, we use the predicted equation
of state, which is in the case of the hard core matrix
\begin{equation}
  \beta\mu_1 = \ln(\eta_1 \Lambda_1/\sigma_1) -
  \ln(1-\eta_0-\eta_1) + \frac{\eta_1+s \eta_0}{1-\eta_0-\eta_1},
\end{equation}
which we set equal to the chemical potential in the reservoir,
\eqref{EQreservoirEqos}. For small $\eta_1^r$ this gives the exact
result, \eqref{EQkHardCoreLowDensity}. In the high density limit,
however, we do not recover \eqref{EQkHardCoreHighDensity}, but obtain
$K(\eta_1^r\to 1)=1-\eta_0$. Note that the small $\eta_0$-expansion of
the exact result, \eqref{EQkHardCoreHighDensity}, is
$K(\eta_1=1)=1-(1+s/2)\eta_0 + O(\eta_0^2)$, differing already in
linear oder.  For intermediate values of $\eta_1^r$ we obtain
numerical solutions. 

\begin{figure}[p]
\includegraphics[width=0.4\columnwidth,angle=-90]{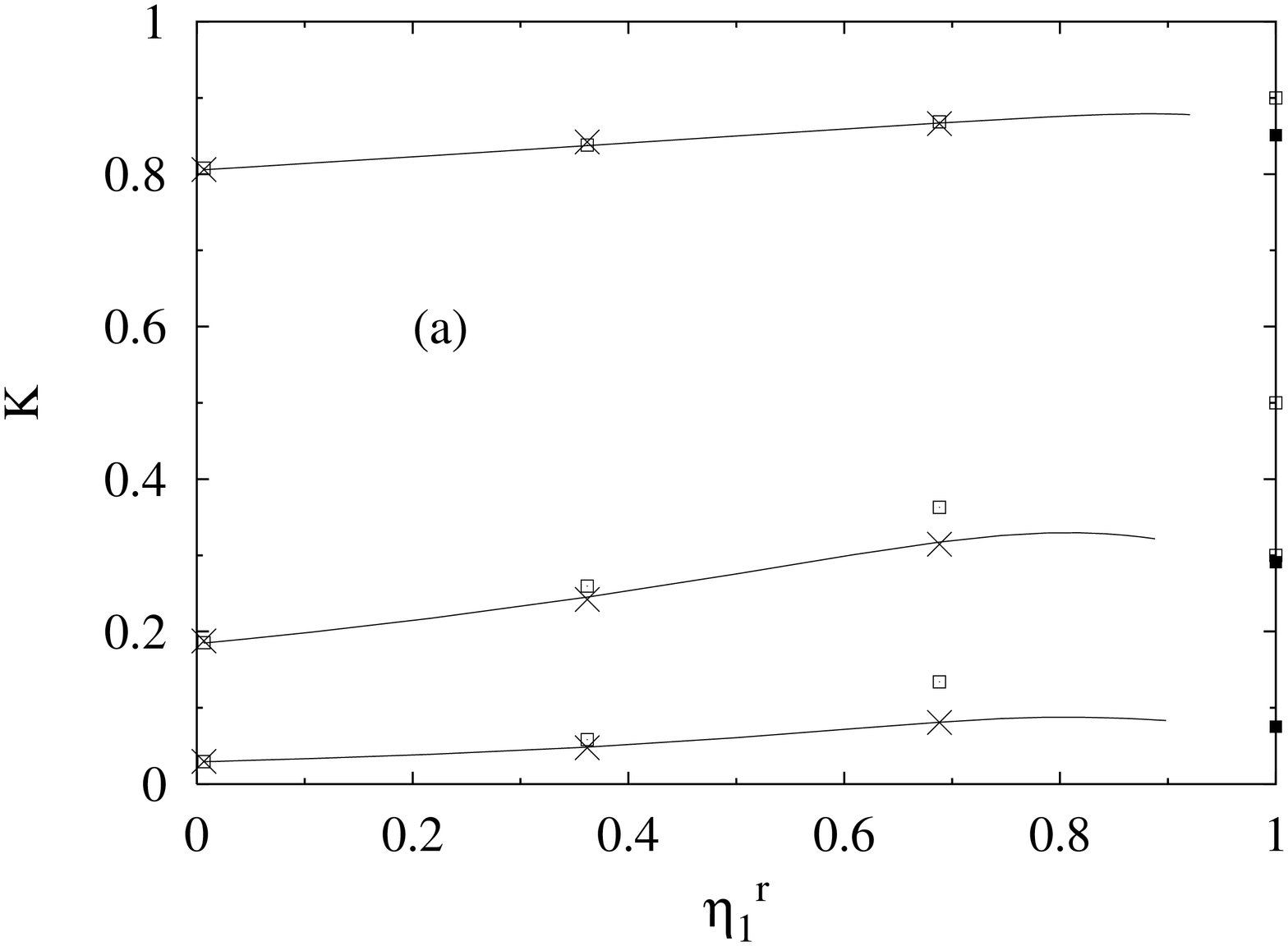}
\includegraphics[width=0.4\columnwidth,angle=-90]{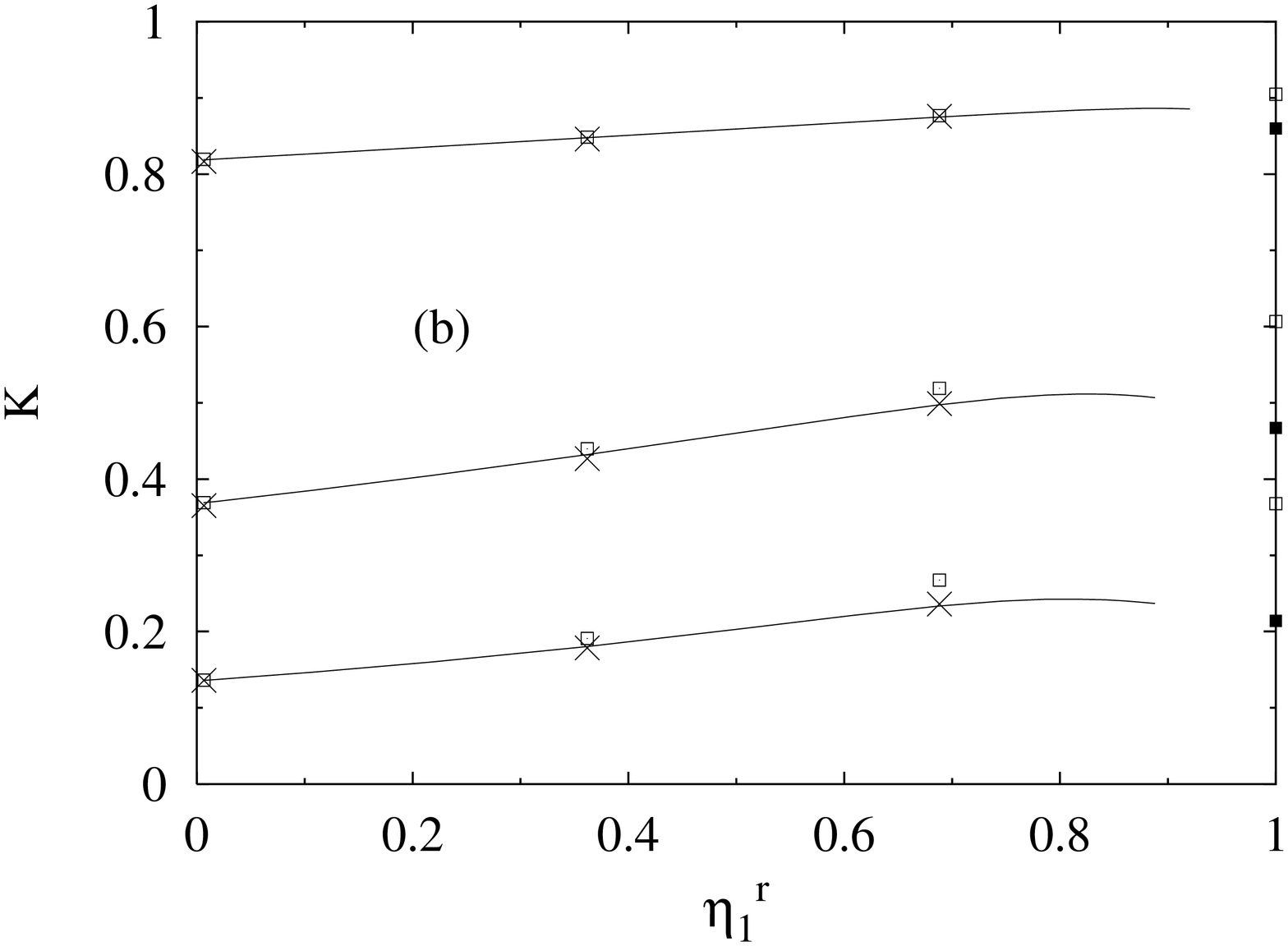}
\caption{Partition coefficient $K=\eta_1/\eta_1^r$ for 1d hard rods
  immersed in a 1d matrix of quenched rods as a function of the
  packing fraction in a reservoir of rods, $\eta_1^r$. Shown are
  results from the elementary calculation, \eqref{EQkFinal}, (lines
  and full squares), DFT with explicit matrix averaging (crosses) and
  QA DFT (open squares).  a) Matrix particles are interacting with a
  hard core potential and possess packing fraction
  $\eta_0=0.1,0.5,0.7$ (from top to bottom). b) Matrix particles are
  ideal and posses packing fraction $\eta_0=0.1,0.5,1$ (from top to
  bottom).}
\label{FIGpartitionCoefficient}
\end{figure}

Results for $K$ as a function of reservoir packing fraction,
$\eta_1^r$, are displayed in Fig.\ \ref{FIGpartitionCoefficient}a for
three different matrix packing fractions, $\eta_0=0.1,0.5,0.7$. For
the lowest value considered, $\eta_0=0.1$, $K$ slightly increases as a
function of $\eta_1^r$. Remarkably a maximum is reached at
$\eta_1^r\sim 0.9$, significantly smaller than the close-packing
limit, $\eta_1^r=1$. We could not obtain numerical results for
$0.9<\eta_1^r<1$, but the available data clearly tend towards the
limiting value, obtained through \eqref{EQkHardCoreHighDensity}. For
the intermediate value $\eta_0=0.5$ the variation of $K$ with
$\eta_1^r$ is stronger and the maximum is more pronounced. For
$\eta_1^r=0.7$ hardly any adsorbate particles can enter the matrix and
the resulting values are $K\lesssim 0.1$, the variation with
$\eta_1^r$ being similar to the above cases. The exact results plotted
in Fig.\ \ref{FIGpartitionCoefficient} are obtained from the
elementary calculation above (lines) and from the DFT with explicit
matrix averaging described in Sec.\ \ref{SECequilibriumDFT}. Both
agree with high numerical accuracy. The result from the QA DFT is
exact in the low-density limit and stays accurate for intermediate
$\eta_1^r$. At $\eta_1^r\sim 0.7$ deviations emerge, overestimating
the values of $K$. Moreover, the maximum is absent, and monotonic
increase with $\eta_1^r$ is found. We hence conclude that the overall
performance of the QA DFT is very satisfactory, but that the
intriguing non-monotonic behavior near close-packing is missed.

The situation in the case of the ideal matrix is similar. The equation
of state from QA DFT is
\begin{equation}
  \beta\mu_1 = \ln(\eta_1 \Lambda_1/\sigma_1)
  -\ln\left(\e^{-\eta_0}-\eta_1\right) +
  \frac{s \eta_0 \e^{-\eta_0}+\eta_1}{\e^{-\eta_0}-\eta_1}.
\end{equation}
Again the result from QA DFT for $\eta_1^r\to 0$ equals the exact
result, \eqref{EQkIdealHighDensity}. In the limit $\eta_1^r\to 1$ the
result from QA DFT is $K(\eta_1^r\to 1)=\exp(-\eta_0)$, which
overestimates the exact expression, \eqref{EQkIdealHighDensity}.  See
Fig.\ \ref{FIGpartitionCoefficient} b) for numerical results for $K$
as a function of $\eta_1^r$ for matrix packing fractions
$\eta_0=0.1,0.5,1$. Compared to the case of the hard core matrix, at
equal densities $K$ is larger, clearly due to the more open void
structure of the ideal matrix particles. Again $K$ displays a maximum
near $\eta_1^r\sim 0.8$, which is not captured within the QA
DFT. Nevertheless, the overall agreement is again very reasonable, and
gives us confidence to turn to inhomogeneous situations.

\subsection{Behavior at the matrix surface}
\begin{figure}
\includegraphics[width=.3\columnwidth,angle=-90]{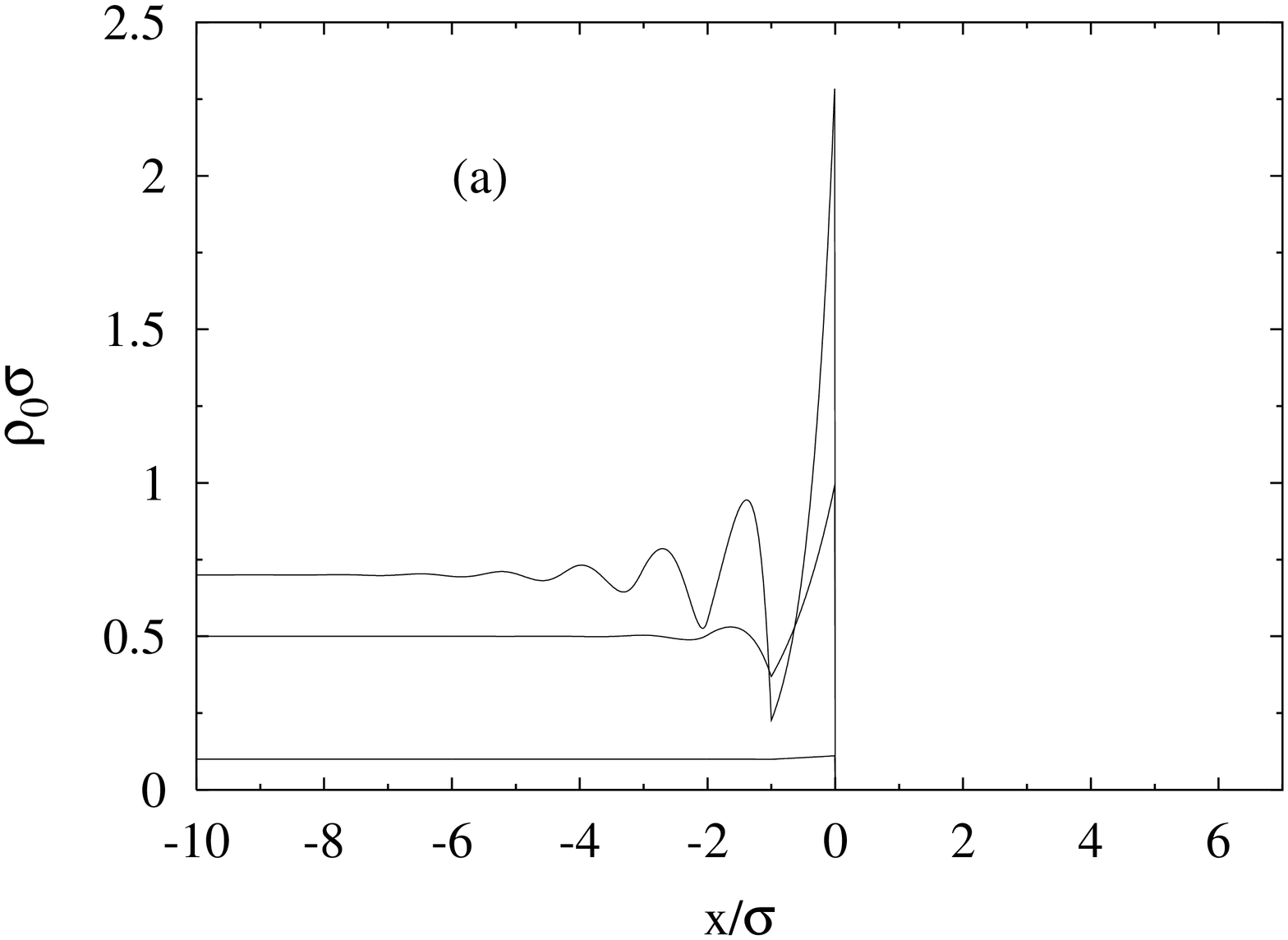}
\includegraphics[width=.3\columnwidth,angle=-90]{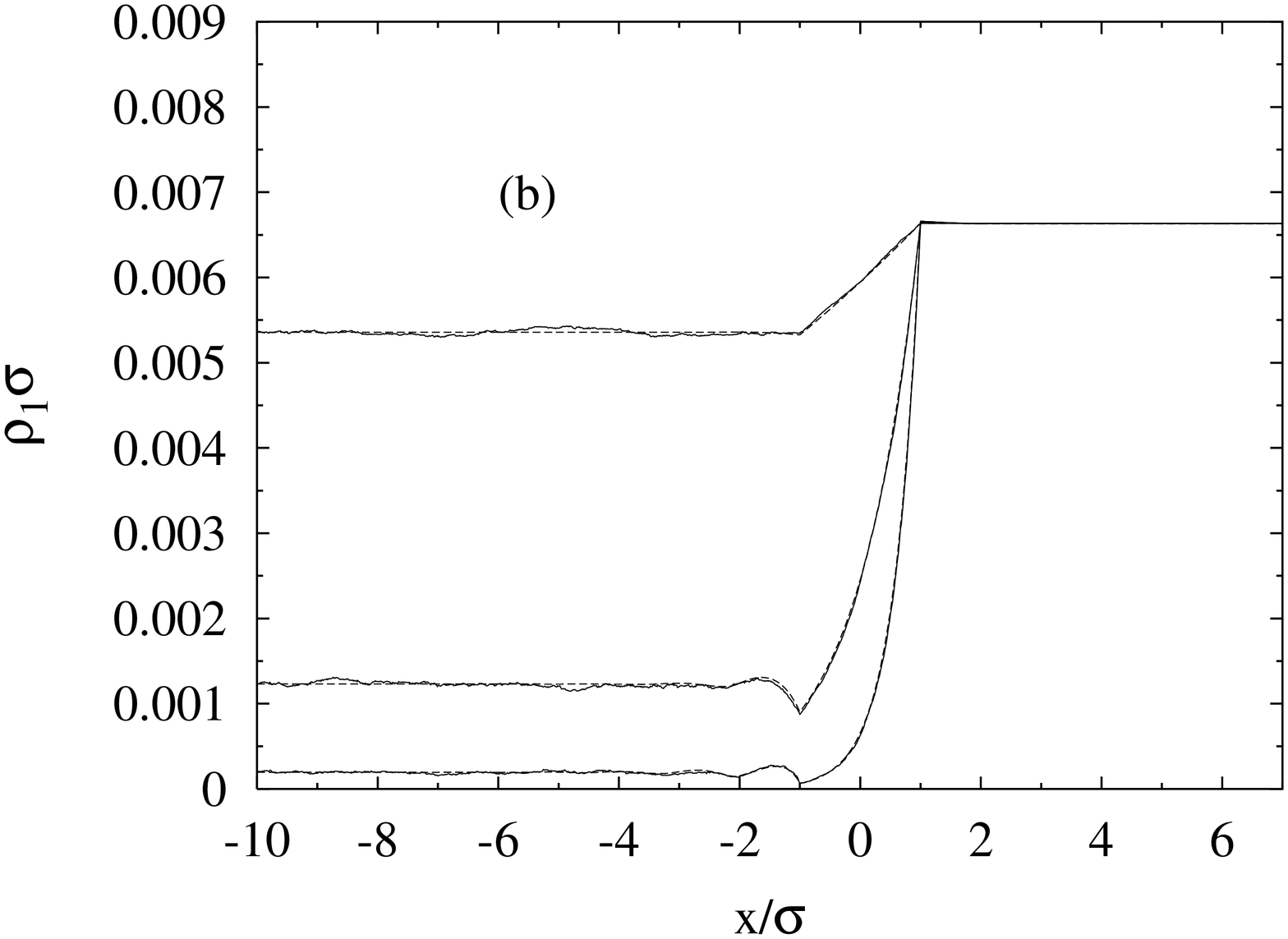}
\includegraphics[width=.3\columnwidth,angle=-90]{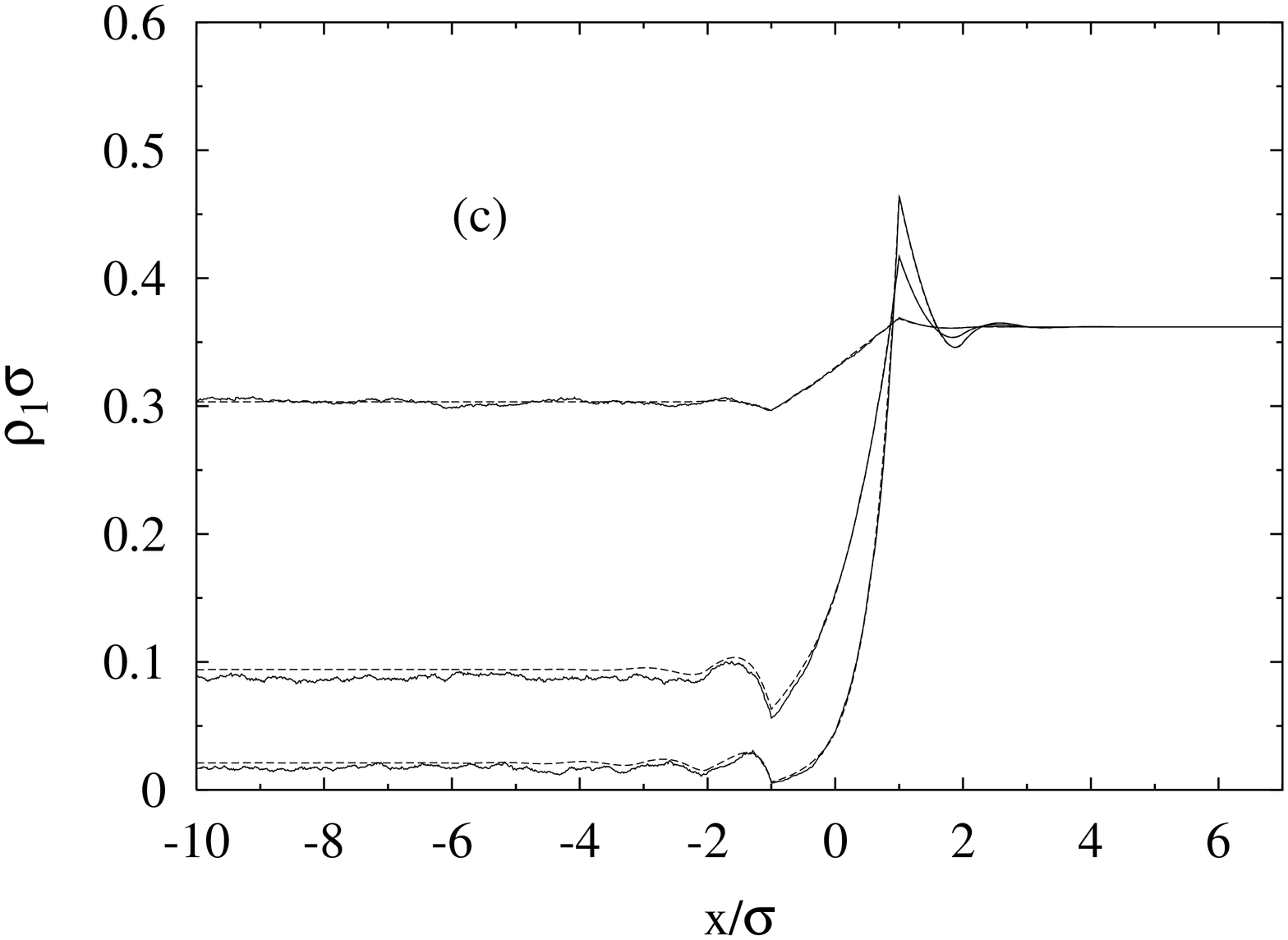}
\includegraphics[width=.3\columnwidth,angle=-90]{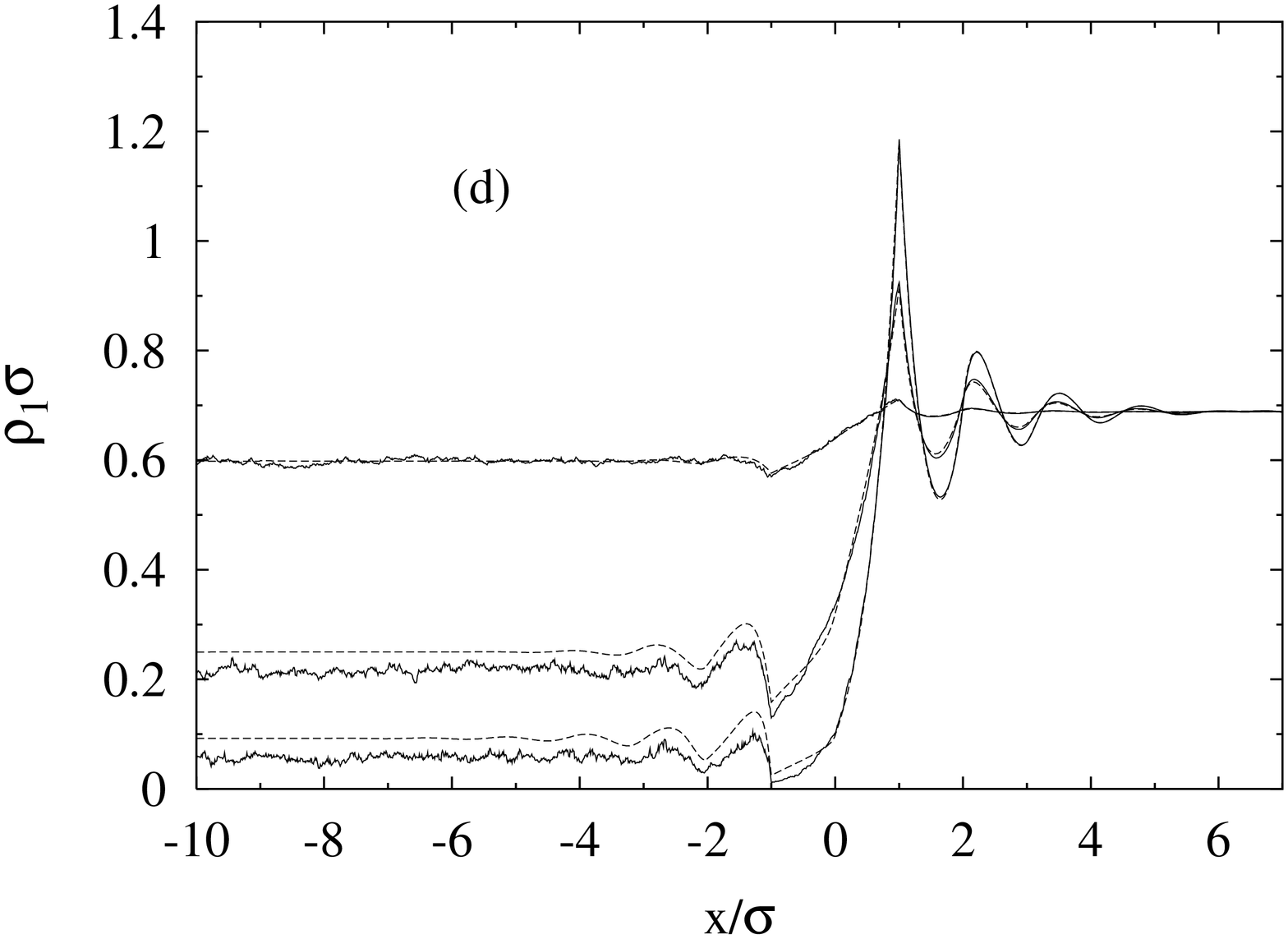}
\caption{Behavior of adsorbate particles near the surface of a matrix
  of quenched rods immersed in a hard core matrix. a) Density profile
  of the matrix particles, $\sigma\rho_0(x)$, as a function of the
  scaled distance $x/\sigma$ for packing fractions
  $\eta_0=0.1,0.5,0.7$ (from bottom to top). Adsorbate density
  profiles $\sigma\rho_1(x)$ are shown for $\eta_0$ (from top to
  bottom) for $\beta\mu_1=-5$ (b), 0 (c), and 3 (d). Results from the
  QA DFT (dashed lines) are compared with those of the exact treatment
  (full lines).}
\label{FIGrhozOneHardCore}
\end{figure}

We consider matrix distributions that are generated by a hard wall,
described by the external potential
\begin{equation}
  \phi_0^\ext(x)=\begin{cases} 0 & x>0 \\ \infty & \rm
  otherwise, \end{cases}
  \label{EQhardWall}
\end{equation}
acting before the quench. In order to obtain $\rho_0(x)$, we solve the
minimization condition, \eqref{EQminZero}, using Percus' functional,
given through \eqref{EQfexcOneComponent}-\eqref{EQphiZeroDimHC}, as
$\F_0^\exc$ in \eqref{EQomegaZeroQA}. This provides an exact
(numerical) solution, see Fig.\ \ref{FIGrhozOneHardCore}a) for matrix
density profiles for three different matrix packing fractions
$\eta_0=0.1,0.5,0.7$. As $\eta_0$ increases the contact value at the
wall increases and the layering near the wall becomes more pronounced,
i.e.\ the amplitude of the density oscillations grows.

The density profile of the adsorbate being exposed to such an
inhomogeneous matrix (note that the hard wall, \eqref{EQhardWall},
only acts on the matrix particles) are obtained using either the
explicit matrix-averaging procedure of Sec.\ \ref{SECequilibriumDFT}
or the QA DFT of Sec.\ \ref{SECquenchedAnnealedDFT}. In Fig.\
\ref{FIGrhozOneHardCore}b-d we display results for $\rho_1(x)$ for
$\eta_0=0.1,0.5,0.7$. For $x>0$ far away from the surface, the
reservoir value is reached, $\rho_1(x\to\infty)=\rho_1^r$; for $x<0$
deep inside the matrix, the asymptotic value is $\rho_1(x\to-\infty)=K
\rho_1^r$, as studied above. For the lowest adsorbate density
considered, result for $\beta\mu_1=-5$ are displayed in Fig.\
\ref{FIGrhozOneHardCore}, the crossover between the limiting cases
happens in a narrow interval $-\sigma<x<\sigma$. Already for
$x>\sigma$ the profile is flat. Inside the matrix, however, for
$x<\sigma$ there are small oscillations with wavelength of the order
of $\sigma$ that decay rapidly with increasing distance from the
surface. These oscillations are not due to the correlations between
adsorbate particles, but are merely ``imprinted'' by the inhomogeneous
{\em matrix} profiles as shown in Fig.\ \ref{FIGrhozOneHardCore}a.
Increasing the adsorbate chemical potential, see Fig.\
\ref{FIGrhozOneHardCore}c for $\beta\mu_1=0$, leads to stronger
structuring outside the matrix, $x>\sigma$. This layering is similar
to that at a hard wall, but with significantly smaller
amplitude. Clearly, this ``washing out'' is due to the average over
the disorder. Note that for a given matrix configuration $\{x_i\}$ the
matrix particle closest to the surface, say $x_{N_0}$, exerts a hard
interaction on the fluid particles with $X_j>x_{N_0}$, and indeed the
resulting $\rho_1(x)$ is that of a hard wall. The disorder-average
then leads to the ``washing out''. For $\beta\mu_1=3$, shown in Fig.\
\ref{FIGrhozOneHardCore}d, even stronger layering is observed. As
expected from the bulk analysis above, the QA DFT overestimates $K$
and hence the adsorbate density inside the matrix, which is clearly
visible for $x<-\sigma$.  The shape of the curves, however, is
predicted very accurately. Moreover, for $x>-\sigma$, the QA DFT lies
practically on top of the exact result.

\begin{figure}
\includegraphics[width=.3\columnwidth,angle=-90]{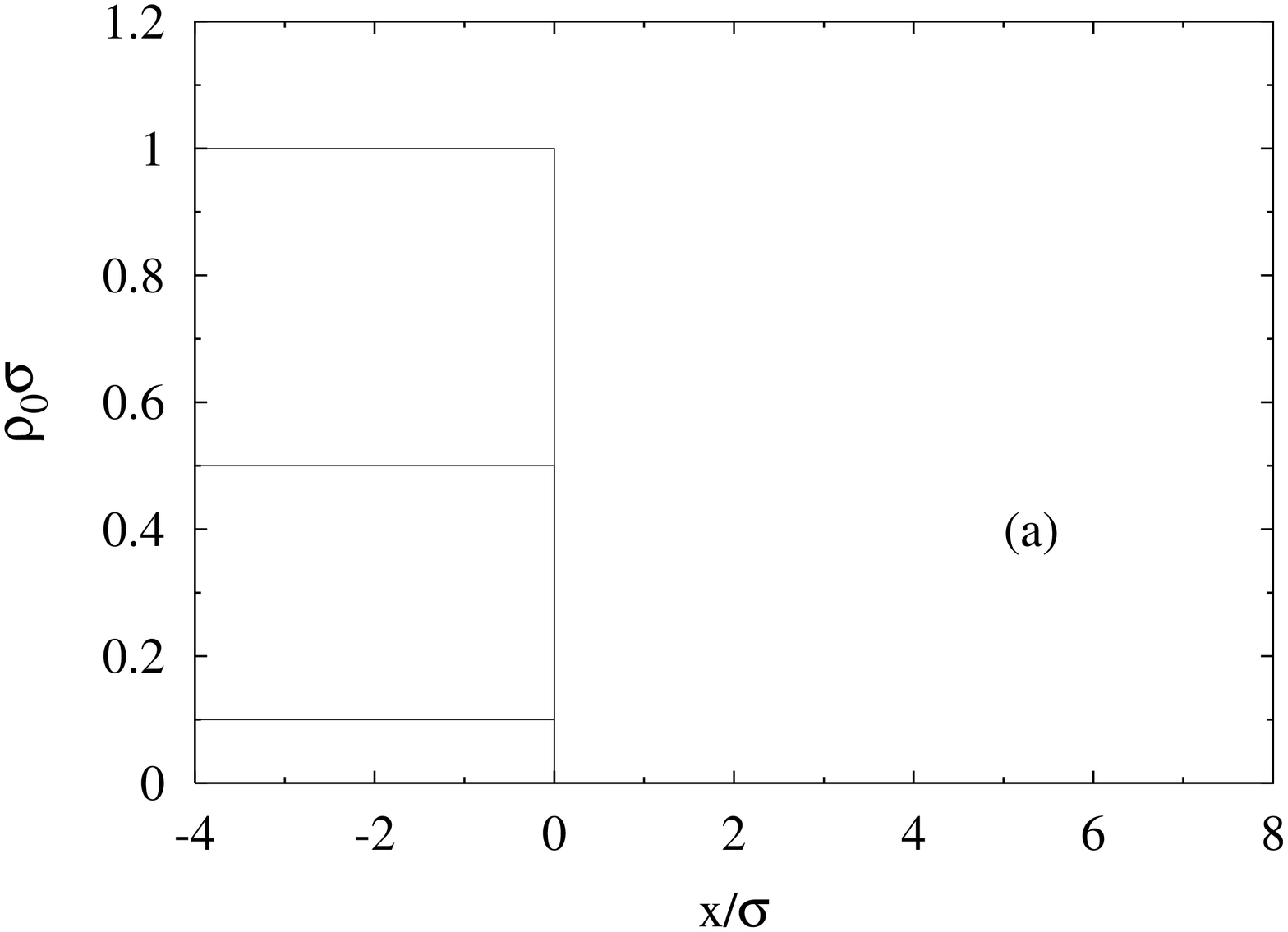}
\includegraphics[width=.3\columnwidth,angle=-90]{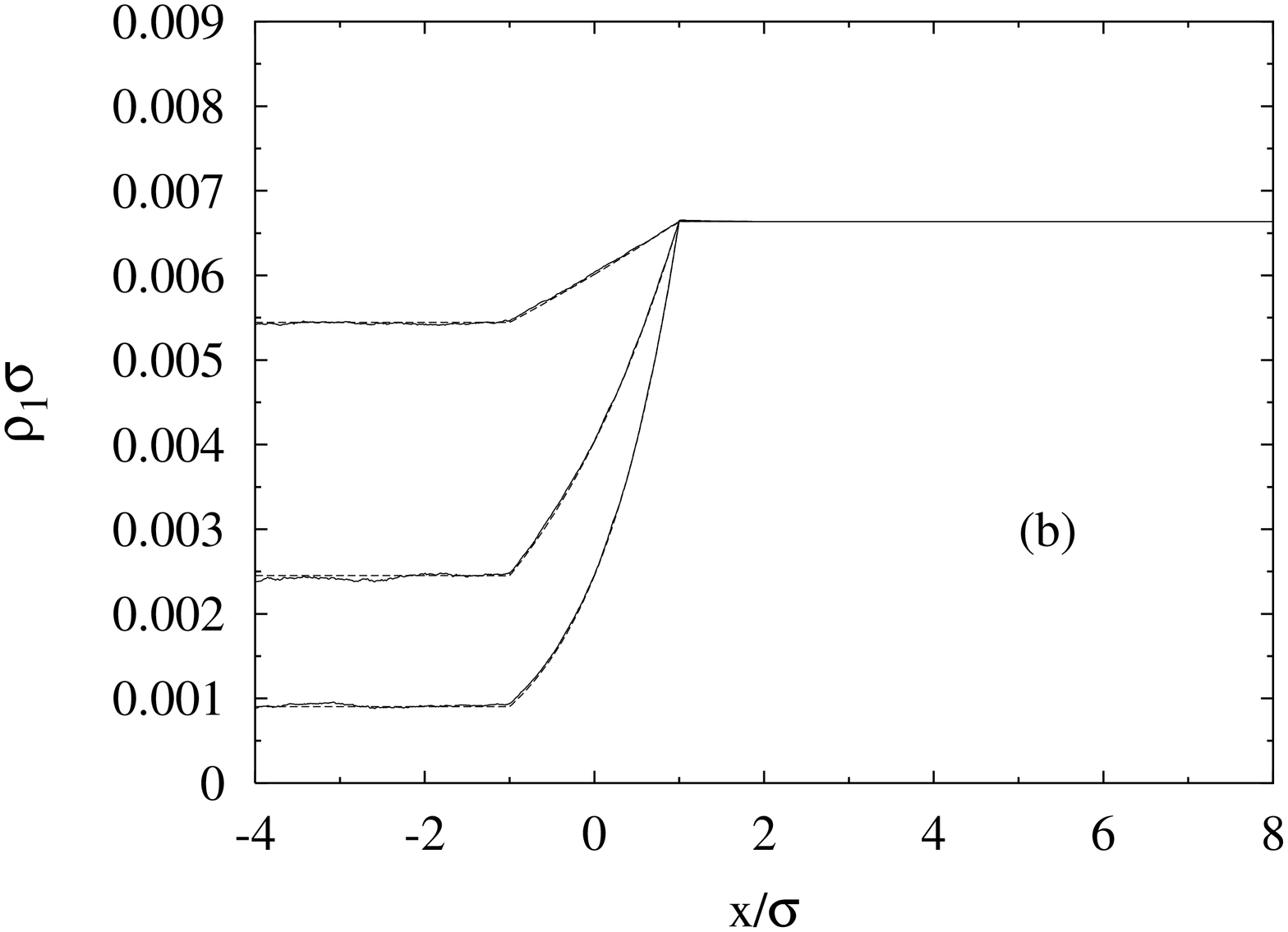}
\includegraphics[width=.3\columnwidth,angle=-90]{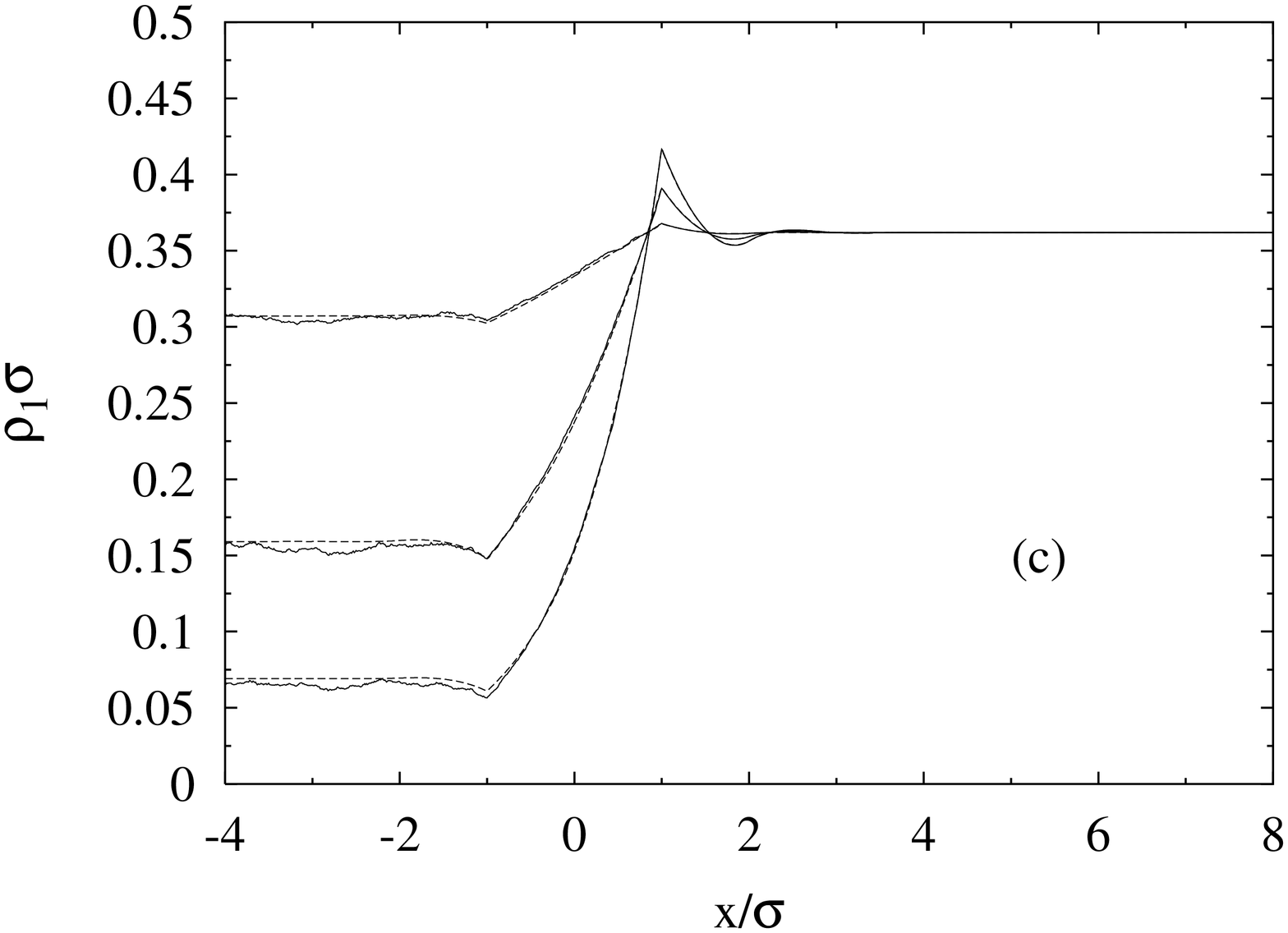}
\includegraphics[width=.3\columnwidth,angle=-90]{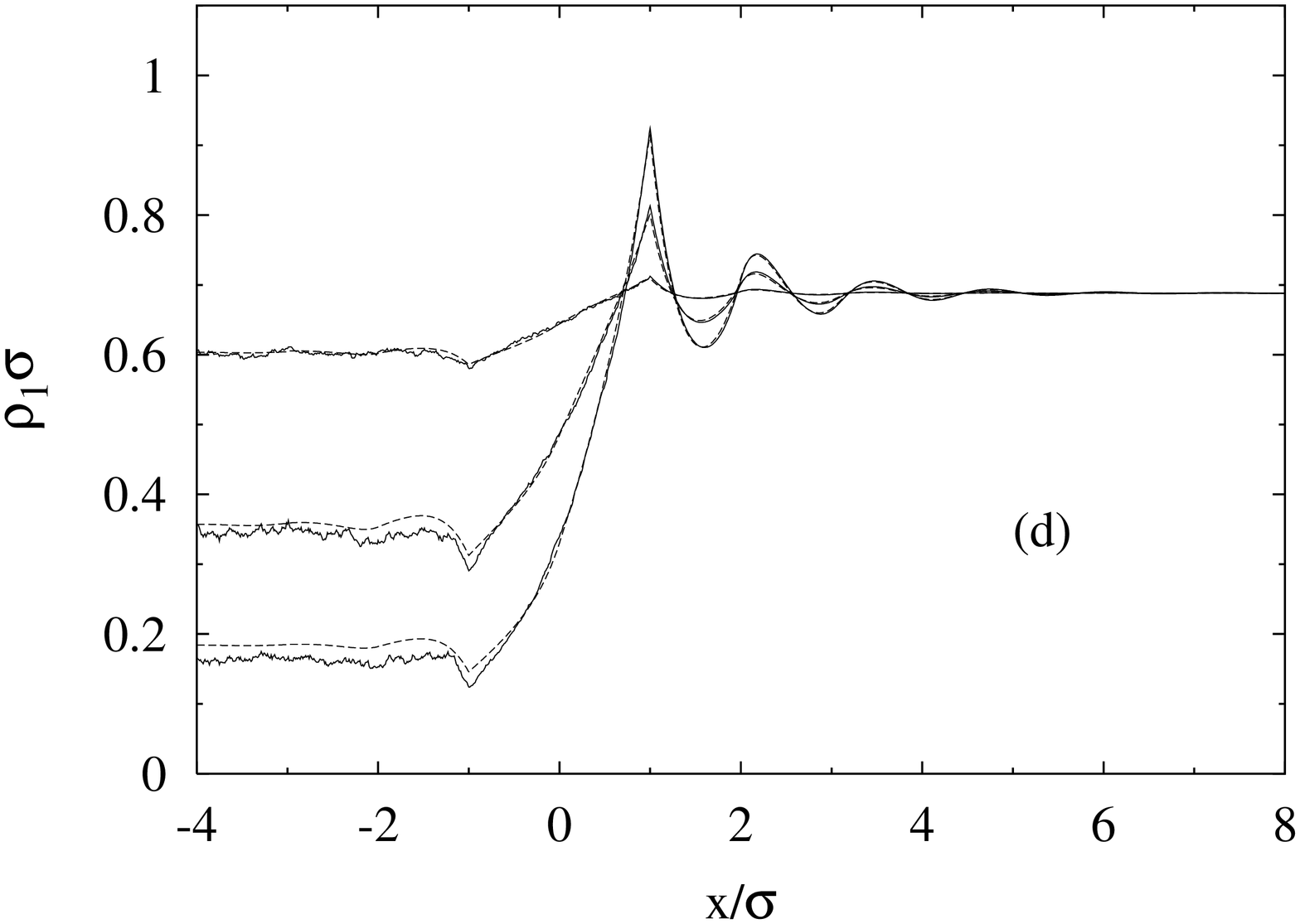}
\caption{Same as Fig.\ \ref{FIGrhozOneHardCore}, but for ideal matrix
  particles of packing fraction $\eta_0=0.1,0.5,1$.}
\label{FIGrhozOneIdealMatrix}
\end{figure}

In the case of the ideal matrix particles exposed to the hard wall,
Eq.\ \eqref{EQhardWall}, leads to step function density profiles, see
Fig.\ \ref{FIGrhozOneIdealMatrix}a, hence is very different from
highly structured profile of the hard core matrix, Fig.\
\ref{FIGrhozOneHardCore}. The general trends upon varying the matrix
density and the adsorbate chemical potential, see Fig.\
\ref{FIGrhozOneIdealMatrix}b-d for results for the same values of
$\beta\mu_1$ as above, is similar. The ideal matrix allows to consider
higher matrix packing fractions, and we have gone up to $\eta_0=1$. It
is to be noted that oscillations do appear for $x<\sigma$, although
the matrix profile is uniform. These oscillations clearly arise from
packing effects between adsorbate particles. Again the results from
the QA DFT agree very well with those of the exact calculation.

\section{Conclusions}
\label{SECconlusions}
In conclusion we have applied a recent DFT for adsorbate fluids in
random matrices to the one-dimensional hard core model. Comparing with
analytic and numerical exact solutions in bulk and at matrix
interfaces we have tested the accuracy of the QA DFT. The general
performance is remarkable, both for the bulk partition coefficient and
for the inhomogeneous density profiles. Subtleties like non-monotonic
variation of the partition coefficient upon increasing adsorbate
density are not captured by the QA DFT.

We note that by confining mesoscopic colloidal particles in narrow
channels \cite{wei00} one-dimensional model fluids are experimentally
accessible. In principle we could imagine modifying setups of
confining groves as described in Ref.\ \cite{wei00} in order to
prepare model systems that resemble the model described in the current
work.  Possible future work could be devoted to the impact of quenched
disorder on three-dimensional narrow channels
\cite{goulding00,goulding01long}.  Furthermore, it would be
interesting to study the asymptotic decay of correlation functions
\cite{evans93,evans94} of fluids with quenched disorder more
systematically. Finally, whether non-monotonic variation of the
partition coefficient appears in 3d hard core models is an intriguing
question.


\vspace{11mm} We thank Hartmut L\"owen, Jos\'e A. Cuesta, Ren\'e van
Roij, and Martin-Luc Rosinberg for very useful comments.
This research is supported by the SFB TR6 ``Physics of colloidal
dispersions in external fields'' of the DFG.  The work of MS is part
of the research program of the {\em Stichting voor Fundamenteel
Onderzoek der Materie} (FOM), that is financially supported by the
{\em Nederlandse Organisatie voor Wetenschappelijk Onderzoek} (NWO).

\bibliographystyle{prsty} 
\bibliography{FMF}

\clearpage

\clearpage

\end{document}